\documentclass[lettersize,journal]{IEEEtran}

\usepackage{amsmath,amsfonts}
\usepackage{algorithmic}
\usepackage{algorithm}
\usepackage{graphicx}
\usepackage{textcomp}
\usepackage{threeparttable}
\usepackage{xcolor}
\usepackage{multirow}
\usepackage{pifont}

\newcommand{\revision}[1]{\textcolor{black}{#1}}
\newcommand{\journal}[1]{\textcolor{black}{#1}}
\newcommand{\tcad}[1]{\textcolor{black}{#1}}

\begin{document}
	
	
	\title{Enhancing IoT Malware Detection through Adaptive Model Parallelism and Resource Optimization}
	
	\author{Sreenitha Kasarapu,~\IEEEmembership{Student Member,~IEEE}, Sanket Shukla, ~\IEEEmembership{Student Member,~IEEE}, \\  Sai Manoj Pudukotai Dinakarrao,~\IEEEmembership{Member,~IEEE} \\
		\thanks{S. Kasarapu, S. Shukla, and S. M. P. Dinakarrao are associated with the Department of Electrical and Computer Engineering, George Mason University, Fairfax, VA, 22030, USA. Email: \{skasarap,sshukla4,spudukot\}@gmu.edu}
		\thanks{This work was supported by the Commonwealth Cyber Initiative, and investment in the advancement of cyber R\&D innovation, and workforce development. For more information about CCI, visit www.cyberinitiative.org}
	}

	\maketitle

	\begin{abstract}

		The widespread integration of IoT devices has greatly improved connectivity and computational capabilities, facilitating seamless communication across networks. Despite their global deployment, IoT devices are frequently targeted for security breaches due to inherent vulnerabilities. Among these threats, malware poses a significant risk to IoT devices. The lack of built-in security features and limited resources present challenges for implementing effective malware detection techniques on IoT devices. Moreover, existing methods assume access to all device resources for malware detection, which is often not feasible for IoT devices deployed in critical real-world scenarios. To overcome this challenge, this study introduces a novel approach to malware detection tailored for IoT devices, leveraging resource and workload awareness inspired by model parallelism. Initially, the device assesses available resources for malware detection using a lightweight regression model. Based on resource availability, ongoing workload, and communication costs, the malware detection task is dynamically allocated either on-device or offloaded to neighboring IoT nodes with sufficient resources. To uphold data integrity and user privacy, instead of transferring the entire malware detection task, the classifier is divided and distributed across multiple nodes, then integrated at the parent node for detection. Experimental results demonstrate that this proposed technique achieves a significant speedup of 9.8 $\times$ compared to on-device inference, while maintaining a high malware detection accuracy of 96.7\%.
	\end{abstract}

	
\section{Introduction}


Recent advancements and innovations in Internet-of-Things (IoT) devices have fueled the growth and extensive deployment of a network comprised of intelligent IoT devices \cite{iot-1}. These devices find application in various domains, including consumer electronics such as smart homes, smart cars, and smart grids, as well as in defense systems \cite{iot-1}. Despite offering numerous benefits, IoT devices and networks have become attractive targets for cyber attackers seeking unauthorized access to user information \cite{Abbas2016BigDI}. Notably, malicious applications, commonly referred to as malware, pose a significant threat to IoT devices, with cyber-attacks often executed through the deployment of such malware \cite{stat_1}. Malware, characterized as malicious software or applications, is designed to infiltrate devices, enabling unauthorized access to sensitive information such as passwords and financial data, and allowing manipulation of stored data without user consent.


Malware stands out as a significant threat, primarily due to its ease of creation and the limited verification capabilities to execute third-party applications on IoT devices \cite{stat_1}. The security risks for IoT networks tend to escalate each year, with an exponential increase observed annually \cite{stat_1}. In 2021 alone, over 5.4 billion malware attacks were recorded, with the first half of 2022 already witnessing 2.8 billion attacks \cite{stat_1}. Adversaries leverage technological advancements to develop sophisticated malware, aiming to evade detection. Records indicate that an average of more than 8 million malware threats are identified daily in recent years \cite{stat_2}.

The significant surge in malware attacks and security threats has heightened concerns regarding the security of IoT devices, potentially hindering their deployability. This underscores the need for techniques capable of detecting malware in IoT devices and mitigating the exploitation of user data. Several studies have been put forth to address malware detection on IoT devices \cite{Wurm'16, Ronen'16}.
However, the existing works primarily suffer from four challenges:

\textbf{(1) Real-time Malware Detection:} 
Detecting malware during runtime with minimal latency is crucial, as malware can have severe consequences and can be challenging to detect once its payload is activated. Recently, two different approaches have emerged for malware detection: static analysis and dynamic analysis \cite{sta_dy, dynamic}. Static analysis involves examining the internal structure of malware binaries without actually executing the binary executable files in a non-runtime environment. On the other hand, dynamic analysis inspects binary applications for malware traces by executing them in a sandbox environment. Unlike static analysis, dynamic analysis is a functionality test, which makes it better at identifying the presence of malware in an application.

Recent works on malware detection (both static and dynamic analysis) 
techniques utilize a variety of Machine Learning (ML) techniques 
to enhance the performance \cite{nataraj}. 
Among the ML-based malware detection techniques, the CNN-based image classification technique \cite{cnn_detect} is observed to be 
efficient due to its prime ability to learn image features.
The emerging trends of malware indicate that the malware developers create advanced malware by employing techniques such as code-obfuscation, metamorphism, and polymorphism \cite{morphism, sanket_dac_2021, sanket_date_2023} to
mutate malware binary executables and modify the static and dynamic application traces (signatures)
and evade malware detection.
This further enhances the complexity of malware detection making the malware detection incur large latency.

\textbf{(2) Reliable Feature Extraction:}
Despite the abundance of research on malware detection \cite{dynamic, cnn_detect}, there is a persistent challenge in reliably extracting input features that contribute to effective malware detection \cite{HPC_FCN}. Regardless of the effectiveness of the underlying analysis technique, whether machine learning (ML) or non-ML, if the extracted features are not reliable, the malware detection task becomes unreliable.
A popular technique to address this challenge is the utilization of hardware performance counters (HPC), device, and network features for node-level malware detection. This approach aims to minimize overheads and meet latency requirements \cite{HPC_FCN}. HPCs can assist in distinguishing between malware and benign applications with low overheads. However, concerns have been raised regarding the reliability of using HPC information for security purposes in recent years \cite{Das'19}. For example, in Intel Pentium 4 processors, the `Instruction count' is often over-counted \cite{Das'19}. Additionally, the coexistence of multiple applications can influence HPC values and trends, leading to non-determinism and unreliability. Therefore, there is a need for improved techniques that can efficiently analyze traits of benign and malware applications while addressing these reliability challenges.

\textbf{(3) Manual Data Acquisition:} 
Supervised learning models are commonly employed for malware detection, utilizing datasets comprising both malware and benign data. However, as the volume of malware data increases annually, there arises a necessity to regularly update these machine learning (ML) models. Yet, the process of collecting, cleaning, and labeling data is labor-intensive. Furthermore, adversaries employ various techniques such as code obfuscation, metamorphism, and polymorphism to enhance the complexity of malware binaries and evade detection \cite{obfuscation, morphism}. In such scenarios, manual data acquisition becomes increasingly challenging. For instance, morphism techniques alter malware binary files to mimic the functionality of standard applications, thereby deceiving the detection capabilities of various methodologies.

Techniques such as code obfuscation \cite{obfuscation} involve encrypting specific sections of code within malware binary files while preserving its functionality. This tactic effectively conceals the presence of malware within embedded systems, exploiting their security vulnerabilities. Another strategy employed to obscure malware identity is stealthy malware \cite{stealthy}, where malware is integrated into benign binaries using randomized obfuscation. Consequently, the benign application exhibits malware-like behavior only after a certain period, rendering it challenging to detect. These sophisticated techniques underscore the complexity of disguising malware and necessitate extensive training to enable machine learning (ML) models to discern hidden malware patterns. Consequently, acquiring the necessary data for training becomes more complex. This highlights the urgency to adopt efficient malware detection techniques that can operate effectively with limited data.


\textbf{(4) Limited Resources on IoT devices:} As previously mentioned, IoT devices are designed with constrained resources to prioritize portability and meet user demands \cite{IOT_ps}. Typically, the bulk of these resources are allocated to executing user applications, with only a limited portion reserved for on-device security measures. Consequently, it is impractical for IoT devices with limited resources to undertake computationally intensive malware detection tasks. Existing approaches either (1) prioritize malware detection at the expense of consuming all available application memory on IoT devices or (2) prioritize user applications, neglecting malware detection capabilities altogether. Both scenarios pose challenges for IoT devices: in the former case, the primary user application's performance is restricted, while in the latter case, user security and privacy are compromised. Thus, there is a pressing need for a technique that can effectively perform malware detection without disrupting the workload of an IoT device.

To address the aforementioned limitations, this work introduces a novel resource-aware and workload-aware model-parallelism-based malware detection technique for IoT devices. This technique enables efficient malware classification without the need for excessive resources from IoT devices. Instead, it employs the distribution of the ML model over neighboring IoT nodes and facilitates malware detection. The application privacy is maintained despite shared resources, as, the model is distributed onto nodes of the same IoT network. The ML model is trained using a few-shot technique to decrease its need for manually annotated image samples. The novel contributions of this work can be outlined as follows: 
\begin{itemize}
	\item This work introduces a methodology for reliably extracting device and network characteristics, laying the foundation for efficient and effective malware detection.
	\item This work implements an automatic assessment of available resources in IoT devices for malware detection. It provides an estimate of whether to offload the malware detection task or not. This analysis is conducted by training a lightweight regressor on the workload of the IoT device and ML model parameters.

	\item The proposed approach involves distributing ML model resources to neighboring devices in a resource-aware manner, taking into account communication and computation overheads for effective malware detection. 
	
	\item We also introduce a code-aware data generation-based few-shot technique aimed at generating mutated training samples to capture the features of actual malware samples. These generated images mimic the complex functionality of malware, addressing the challenge of comprehensive data acquisition.

\end{itemize}

The experimental results prove that the proposed resource-aware model parallelism technique can detect complex malware in IoT networks with an accuracy of over 90\%. 
Experimental analysis shows that the proposed technique can achieve a speed-up of 9.8$\times$ compared to on-device inference while maintaining a malware detection accuracy of 96.7\%.

The rest of the paper is organized as follows: Section \ref{related} describes the related work and its shortcomings and comparison with the proposed model. Section \ref{problem} describes the problem for malware detection in IoT devices. Section \ref{proposed_sol} describes the proposed architecture resource-aware model parallelism, which assists with efficient malware detection in IoT devices, using a distributed runtime model training methodology. The experimental evaluation of the proposed model and comparison with various ML architectures is illustrated in Section \ref{results} and followed by the conclusions drawn from the paper are furnished in Section \ref{conclusion}.

\section{State-of-the-Art}
\label{related}

In this section, we present some of the relevant works proposed in the recent past on malware detection, distributed learning, and few-shot learning. 

\subsection{Malware Detection Techniques}
Malware detection in recent years has gained a lot of interest. We broadly categorize malware detection into two categories. 

\subsubsection{Static Analysis based Malware Detection}

Traditionally, static and dynamic analyses of malware detection are employed. Static analysis \cite{sta_dy} on malware data is performed by comparing the opcode sequences of binary executable files, control flow graphs, and code patterns. This technique is performed in a non-runtime environment, as it doesn't require any executions. 


The work in \cite{nataraj} introduced a technique for malware detection using image processing technique where binary applications are converted into grayscale images. The generated images have identical patterns because of the executable file structural distributions. The paper used the K-Nearest Neighbour ML algorithm for the classification of malware images. Similar approaches include image visualization and classification using machine learning algorithms such as SVM. 
However, these approaches don't address the problem of classifying newer complex malware. 
Neural networks such as artificial neural networks (ANNs) are used extensively to solve the problem of classification, prediction, filtering, optimization, pattern recognition, and function approximation \cite{img_process}, as neurons can capture the features of the images more accurately than other machine learning algorithms. 
However, the fully connected layers of ANN tend to exhaust the computational resources. 
In \cite{cnn_detect, sanket_dhavlle2021novel, sanket_glsvlsi_2022} authors used Convolutional neural networks (CNNs), due to their ability to efficiently handle image data through feature extraction by Convolutional 2D layers and using Maxpooling 2D layers to downsample the input parameters. Thus, serving as an efficient image classification algorithm with lesser resource consumption. 


\subsubsection{Dynamic Analysis based Malware Detection} 

Dynamic analysis is a malware detection technique, performed in a secured runtime environment, like Sandbox. It is a functionality test and the binary files are executed to detect malware functionalities in them. Malware detection using dynamic analysis is performed based on detecting system calls or HPC \cite{dynamic}. Dynamic analysis is much more efficient than static analysis in malware detection. 
Dynamic analysis need a huge amount of resources and is time consuming, so, it is hard to carry on edge devices. 
Furthermore, malware developers implement code obfuscation, metamorphism, and polymorphism \cite{morphism} to mutate malware binary executables. These new strategies in masking malware's identity are stealthy malware \cite{stealthy}, where malware is incorporated into benign applications using random obfuscation techniques. In such cases, dynamic malware analysis produce poor estimations. So there is a need to train these dynamic models with reliable features.

In past, many researchers have leveraged architectural and application features for malware analysis and detection \cite{esweek_18_hh}. In \cite{Bilar_2007} Bilar et al. used the difference of opcodes between known malware and benign as a key to predicting malware. 
However, these proposed techniques require a considerable amount of work to model each program based on instructions. Since the code size increases day by day, modeling programs based on opcodes becomes a time-consuming and computationally expensive task.
Demme et al. \cite{Demme'13} proposed the use of a hardware performance counter (HPC) to monitor
the lower level micro-architectural parameters such as branch-misses, instruction per cycle, and cache miss rate. HPCs can provide access to interior performance information comprehensively with much lower overhead than other methods. In works such as \cite{HPC_ML, sanket_abhijitt_date2021, sanket_cases_2019, sanket_isqed_2024, sanket_rram_glsvlsi_21, sreenitha_kasarapu2021demography, sreenitha_mdpi, sreenitha_sanket_aspdac, sreenitha_sanket_tcad, sreenitha_sanket_glsvlsi, sreenitha_sanket_ubol, sreenitha_sathwika_vlsid, raghul_RS2020, raghul_iscas, raghul_SaravananICDSMLA'19, raghul_SaravananICICNIS'21, raghul_trng2020}, machine learning models like Random Forest, SVM, and Logistic Regressors are used on HPC values obtained at execution, to classify benign and malware classes. In \cite{HPC_FCN}, the authors introduce StealthMiner a novel stealthy malware detection model using time series prediction. They build a Fully Convolutional Neural Network (FCN) on HPC run-time branch instruction features to detect stealthy malware traces.



\subsection{Distributed Learning}

Deep learning has achieved milestone and has valuable applications in cybersecurity, malware detection and other domains. Training deep learning models requires huge amount of time especially with the massive amount of data which needs to be processed. On the other hand, scaling neural network architecture may result in a network with complex parameters, leading to time complexity (i.e. high execution time while training the model). Fortunately, these bottlenecks can be addressed through parallelization paradigms. Parallelization of tasks in deep learning models is one of the
best approaches for accelerating implementation, i.e., it speed-ups the algorithm by minimizing the execution time, allowing complex tasks to be processed with less computational resources and execution time \cite{de2012parallelization}.

Two types of distributed learning techniques are available: data parallelism and model parallelism. In data parallelism \cite{data_parallel}, each node has a copy of the whole ML model which needs to be trained. But, each node is given a different mini-batch of data for training the model. After training the results are collected and combined into an updated model. Though it reduces complexity and inter-node communication, data parallelism suffers from huge memory utilization. Model parallelism \cite{Verbraeken'20} is a technique, where each node has the same data but the ML model is divided. Each node contains only a single layer of the neural network to be trained. Node-to-node communication is done for weight sharing and back-propagation. Model parallelism is suitable to train a massive ML when there are limited resources.    


In \cite{algebraic}, authors propose linear-algebraic-based model parallelism for deep learning networks. This framework allows the parallel distribution of any tensor in the DNN. Model parallelism is also mainly used in natural language processing. In \cite{megatron}, authors train a multi-billion parameter-based transformer language model. With the help of multiple GPU nodes and pipeline structures, they could train such a gigantic model. It also achieves state-of-the-art speedups. In \cite{3d_transformers}, authors build a 3-dimensional distributed model to accelerate the training in the language model. They use a 3D model to complement matrix multiplication and vector operations in the transformer models. \revision{To the best of our knowledge, this is the first work that employs model parallelism for the purpose of malware detection. }



\subsection{Few-Shot Learning}

\journal{With a consistent increase in malware applications each year \cite{stat_1}, there is a constant need to update the ML models involved in malware detection algorithms. But complex data availability and continuous data collection for different cases are difficult. The machine learning and deep learning models need to be updated with each new type of training sample to generalize well. Due to this, the efficiency of machine learning models for malware detection is often debated. So there is a need to build an efficient malware detection model with only a few samples that do not need constant updating. Few-shot learning is a supervised learning technique that aims to learn different class concepts using a few samples. And could improve ML models which have limited complex data availability.}

The important frameworks for few-shot learning are data augmentation techniques. These models improve the feature extraction capability of few-shot learning algorithms. Models such as Generative Adversarial Networks (GAN) \cite{gol}, Variational Autoencoders (VAE) \cite{vae} and Mixture Density Networks (MDN) \cite{mdn} can generate high-quality samples. GANs can produce new samples by loss minimization in the generated samples, and MDN with the help of Gaussian Mixture Models can produce highly probable samples. VAE with its encoder-decoder architecture is said to reconstruct input data efficiently. Works such as \cite{data_aug}, \cite{data_aug_os, raghul_iscas} use techniques such as reflection, translation and augmentation to generative new samples for training. \cite{mem_aug} used a memory augmentation technique for few-shot learning.


\section{Motivation and Problem Formulation}
\label{problem}


With technology advancements, attackers are introducing complex hidden malware, by sneaking them into general applications. 
This is mathematically represented in Equation \eqref{eq1}. 
Even advanced anti-malware software fails to detect these advanced malware families \cite{morphism}.

\begin{equation}
	\centering
	\mathbb{IOT}_{devices} \leftarrow (B \oplus M)
	\label{eq1}
\end{equation}

As represented in Equation \eqref{eq1}, B represents benign and M represents the malware executables for IoT devices.
The target for the malware is the IoT devices, represented as 
$\mathbb{IOT}_{devices}$. 
One can represent the problem of malware detection on IoT devices as follows:


\begin{equation}
	\begin{aligned}
		& & \mathbb{C}(D^n): {X} \rightarrow {Y} \hskip 0.75em \\
		& \text{s.t.} 
		& \mathfrak{M}[\mathbb{C}] < \mathfrak{M}[node]
		\label{eq2}
	\end{aligned}
\end{equation}


\begin{figure*}[htbp]
	\centering
	\includegraphics[width=7in]{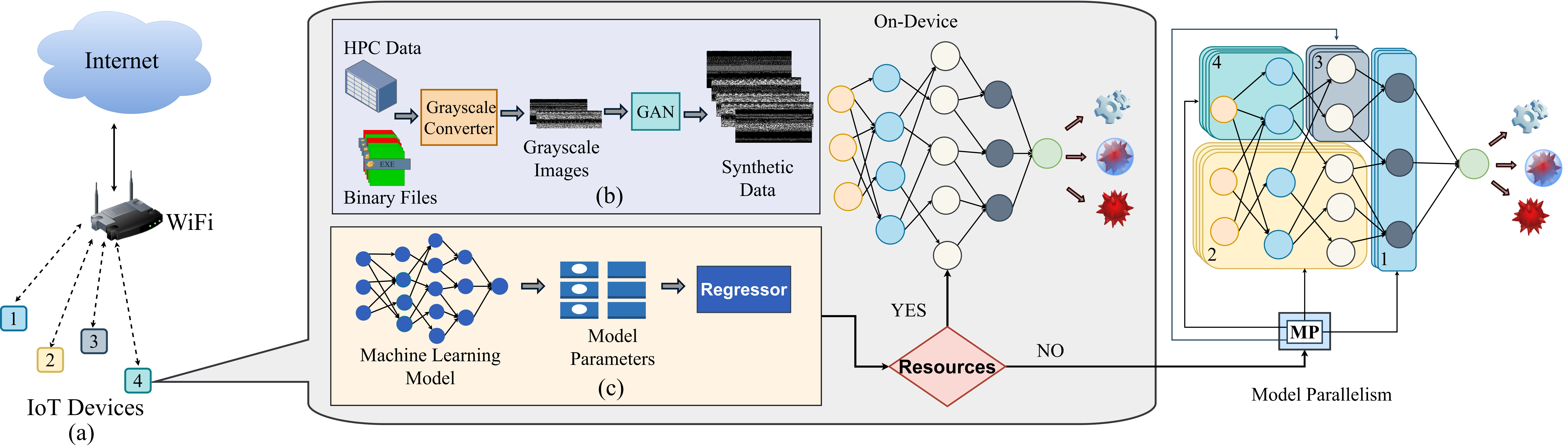}
	\caption{(a) Distributed IoT device framework, (b) HPC and Binary data pre-processing to extract input image dataset and generating additional synthetic samples with Code-Aware Data Generation technique using GANs, (c) Framework to identify the resources in the malware detection model using a lightweight linear regressor} 

\label{fig:overview}
\end{figure*}

\journal{As shown in Equation \eqref{eq2}, $\mathbb{C}$ is a pre-trained classifier trained with dataset $D^n$ to perform malware detection. The dataset $D^n$ contains, a combination of malware $M$ and benign $B$ samples. As a pre-trained model, the classifier $\mathbb{C}$ will not incur any overhead and can be used for inference. This model has the ability to detect if there is malware in any sample $X$ and map it to either malware class ${M}$ or benign class $B$. The output class is represented as $Y$. The memory required to perform inference, represented as $\mathfrak{M}[\mathbb{C}]$ should be less than the available resources in an IoT node, represented 
as $\mathfrak{M}[node]$. If the constraint in equation \eqref{eq2} is not met, 
then the inference task can't be carried out by the device. Also, to produce an effective ML model there is a need for enough training samples $D$. With the need for enough training data and memory, the problem of implementing malware detection in IoT devices can be defined as a dual optimization problem.}

\begin{equation}
\begin{aligned}
\text{maximize} \quad 
& & \sum_{i}\sum_{j} D_{ij}\mathfrak{M}_{ij} \\
\text{s.t.} \quad
& & \sum_{{j\in\mathcal{P}}} d_{xj}=1 \quad\forall i=d,\mathfrak{m}        \\
\label{eq3}
\end{aligned}
\end{equation}

Equation \eqref{eq3} describes the problem of optimizing training data and the available resources such as memory. Our proposed technique solves this by introducing a novel resource-aware malware detection model through off-loading the workload inference to neighboring nodes. We also introduce a code-aware data generation technique to increase the training samples. Thus addressing the problem in IoT devices of limited memory and training data.

\section{Proposed Resource- and Workload-aware Malware Detection}
\label{proposed_sol}

\subsection{Overview of the Proposed Technique}
\revision{The overview of the proposed technique is shown in Figure \ref{fig:overview}. The computations happening at node level are presented. The Figure \ref{fig:overview}(a) represent the IoT devices present in a network. 
	The proposed technique starts with data collection at the IoT device, in which the popular malware and benign application files are collected. 
	Figure \ref{fig:overview}(b) describes the data collection process. 
	The HPC traces are considered as input for the proposed technique to improve the reliability of malware detection. Along with the HPC data,
	the benign binary samples used in IoT devices and malware binary samples which affect the IoT devices are collected. The HPC data and binary files are converted to grayscale images. To increase the training data for better training capabilities synthetic data is generated using code-aware data generation technique is employed. These image samples are fed as input to the machine learning 
	algorithms such as CNNs for effective malware detection. As shown in Figure \ref{fig:overview}(c), an automatic estimation is done using a lightweight regression model to analyze the resources needed to perform the malware detection. Depending on the resource availability, workload in a IoT node and the communication overhead, the malware detection task is either performed on-device or off-loaded to neighbouring nodes with sufficient resources as shown in Figure \ref{fig:overview}. The $MP$ block in Figure \ref{fig:overview} represents the model parallelism task.}

\subsection{Pre-processing and Data
	Collection}

\subsubsection{Generation HPC-based Grayscale Images}





To address the reliability concerns which are not addressed in the existing techniques, 
we propose fine-tuning state-of-the-art model-specific registers (MSRs) available in the modern computing system architectures, which are the source of the HPC information. 
Firstly, to solve the non-determinism challenge in HPCs, we redesign HPC capturing protocols with 
proper context switching and handling performance monitoring interrupt (PMI) units in the system while collecting HPCs. To obtain the HPCs solely for a given application, context switching needs to be accommodated, thereby eliminating the contamination of the 
obtained HPCs. From our preliminary analysis, the overhead (in terms of latency) to perform context switching for MiBench applications is around 3\% of an average application runtime which is 
affordable for enhanced security. Further, to ensure proper context switching and reading of HPCs, PMIs can aid. It has been seen that configuring PMI per process often leads to better capturing of the 
HPCs \cite{Das'19, sanket_iccd_2022, sanket_icmla_2019, sanket_ictai_2019}. Through this two-pronged utilization of context-switching+PMI, we collect reliable HPCs. To address the challenges such as over counting \cite{Das'19}, 
we perform calibration through testing.


We also require the microarchitectural event traces captured through HPCs for 
malware detection. One of the challenges is that there are
a limited number of available on-chip HPCs that one can extract at a given time-instance. However, executing an application generates few tens of microarchitectural events. 
Thus, to perform real-time malware detection, one needs to determine the non-trivial microarchitectural events that could be captured through the limited number of HPCs and yield high detection performance. 
To achieve this, we use principal component analysis (PCA) for feature/event reduction on all the microarchitectural event traces
captured offline by iteratively executing the application. 
Based on the PCA, we determine the most prominent events and monitor them during runtime. The ranking of the events is determined as follows: 
\begin{equation}
	\rho_i = \frac{cov(App_i,Z_i)}{\sqrt{var(App_i) \times var(Z_i)}}
	\label{pca_eq}
\end{equation}
where $\rho_i$ is pearson correlation coefficient of any $i^{th}$ application. $App_i$ is any $i^{th}$ incoming application. $Z_i$ is an
output data contains different classes, backdoor, rootkit, trojan, virus and worm in our case. $cov(App_i, Z_i)$ measures covariance between
input and output. $var(App_i)$ and $var(Z_i)$ measure variance of both input and output data respectively. 
Based on the ranking, we can select most prominent HPCs and monitor them during runtime for efficient malware detection. 
These reduced features collected at runtime are provided as input to ML classifiers which determine the malware class label (${\widehat{Y}}$ $\Rightarrow$ Backdoor, Rootkit, Trojan, Virus and Worm) with higher confidence. 

To the best of our knowledge, this is the first work which captures functionality of dynamic HPC attributes (values) and converts/represents them into grayscale images. We execute malware and benign application in a sandbox environment and capture range of HPC values (e.g. for 20 ns, 40 ns) using Quick HPC tool. Capturing the range of HPC values for a particular executable (benign or malware), illustrates the trend in variation in the HPC values for benign and malware samples. Hence, we have unique patterns in grayscale image for each executable file. However, it should be noted that the grayscale images of same class of malware tend to show similar texture in some portion of the grayscale image. Moreover, the advantage of this technique is the malware payload which is triggered by stealthy and code obfuscated malware can be identified and classified based on HPC based grayscale images because the grayscale texture of triggered malware tend to match either of a  malicious pattern from the generated training data. 



\begin{table*}[ht!]
	\centering
	\caption{Parameter Estimations per Each Layer in a CNN Algorithm }
	\label{tab1}
	\scalebox{0.75}
	{
		\begin{tabular}{|c|c|c|}
			\hline
			
			\textbf{Layers} & \textbf{Description} & \textbf{Parameters} \\
			\hline
			Input  & No learnable parameters  & 0  \\ 
			\hline
			CONV  & (width of filter * height of filter * No. of filters in previous layer+1) * No. of filters in current layer & $f_{conv} = (w*h*p) + 1)*c$   \\ 
			\hline
			POOL & No learnable parameters & 0  \\ 
			\hline
			FC & (current layer neurons * previous layer neurons)+1 * current layer neurons & $f_{FC} = (n_c * n_p) + 1*n_c$ \\ 
			\hline
			Softmax & (current layer neurons * previous layer neurons) + 1 * current layer neurons & $f_{S} = (n_c * n_p) + 1*n_c$ \\ 
			
			\hline
		\end{tabular}
	}
\end{table*}

\subsubsection{Code-Aware Data Generation}
Code-aware data generation technique is a novel data augmentation technique to generate reliable synthetic data. This synthetic data helps in feature extraction from limited samples. The data generation is done using generative adversarial networks (GANs). \journal{It is code-aware because GANs are trained with images constructed from binary code files. So the feature extraction carried out in GANs can be interpreted as capturing the malware code patterns. So in the case of varied test data, there won't be the need to train ML models again. The obfuscated and morphic malware samples, which have hidden malware code blocks can be detected easily as GAN is made able to detect these hidden patterns. This makes the data generation process code-aware.} In the case of HPC samples, grayscale images are constructed based on the functional attributed. GANs are trained with dynamic HPC grayscale images, to generate augmented HPC samples. 
The generated images are loss-controlled which makes them effective in capturing the features of limited available data. GAN consists of two parts a generator and a discriminator. Generator considers a random uniform distribution as a reference to generate new data points. Based on this uniform distribution and input data, generator tries to generate a correlated sample. This generated sample augments the real image with the help of uniform distribution so that when given to a ML model the feature extraction rate improves. The discriminator block of a GAN tries to classify the generated image as real or fake.

The  generator and discriminator are loss-controlled, so that the generator can generate realistic images which are as close to the real images. And the discriminator is trained to invalidate the fake images. This helps the generator to learn and improve its ability to generate data. And discriminator is trained to classify them better.


\begin{algorithm}[htp!]
	\caption{Code-Aware Data Generation Algorithm}
	\label{algo4}
	\begin{algorithmic}[1]

		\STATE {\textbf{Input}}: $D_{l}$ (Dataset with limited data version), $B$ (Benign grayscale images), $M$ (Malware grayscale images), $M_O$ (Random obfuscated malware), $M_{ST}$ (Generated Stealthy malware),\\
		$ D_{l} = \{B + M + M_O + M_{ST}\} $


		\STATE \hskip 1em \textbf{define} CAD\_generator(X):
		\STATE \hskip 2em \textbf{for} {$ X \leftarrow D_{l}$}: \textbf{do} 
		
		\STATE \hskip 3em \textbf{for} {$epoch \leftarrow range(1000)$}: \textbf{do}

		\STATE \hskip 4em $G\_model = define\_generator()$
		\STATE \hskip 4em $D\_model = define\_discriminator()$
		\STATE \hskip 4em $noise \leftarrow vector(256, None)$ 
		\STATE \hskip 4em $X\_{fake} \leftarrow G\_model(noise) $
		
		
		
		\STATE \hskip 4em\textbf{end for}
		\STATE \hskip 3em $D_{mu} \leftarrow G\_model\cdot predict(vector)$
		\STATE \hskip 3em \textbf{end for}
		\STATE \hskip 3em \textbf{return} $D_{mu}$

		\STATE {\textbf{Output}}: $D_{mu}$ (Generated dataset with mutated samples)

	\end{algorithmic}
\end{algorithm}

Algorithm \ref{algo4} takes in the limited version dataset $D_l$ as input. For each class in the dataset, the CAD\_generator(X) function trains a generator and a discriminator. We train our GAN for 1000 epochs (Line 4), enough times to minimize the loss and generate images similar to training data. As represented in the algorithm (Line 5- Line 6), the generator model is described as $G\_model$, and the discriminator model is described as $D\_model$. They are convolutional neural networks, where, $G\_model$ is trained to generate an image when a latent space is given as input. As represented in the algorithm \ref{algo4} (Line 7- Line 9), when a latent noise generated by function $vector()$ of size $(256, None)$ is 
given as input, it generates an image of size $(32,32)$. The $D\_model$ tries to classify the generated fake image X\_{fake}. A loss function is generated for $D\_model$ and $G\_model$. To decrease the gradient loss, the generator learns to generate better fake images $X\_{fake}$, and the discriminator keeps on learning to classify them. After 1000 epochs, the generator model learns enough to be able to generate realistic fake images. 
So vectors of latent spaces are created to generate mutated data by using the $model\cdot predict()$ function, they are represented using dataset $D_{mu}$ (Line 12). 

\revision{
	\vspace{-0.5 em}
	\begin{equation}
		\centering
		D_{mu}(X) \sim D_w(X) 
		\vspace{-1em}
		\label{eq5}
\end{equation}}

The samples in the generated synthetic dataset represented as $D_{mu}(X)$ have a high correlation with real samples $D_w(X)$. A few shots of real samples $D_w(X)$ are used for training a CNN classifier along with the generated synthetic data $D_{mu}(X)$ for malware detection.

\journal{After the data generation happens, a dataset is built using a few shots of real data, and a CNN model is trained with this data. The augmented data generated in the code-aware data generation technique helps in training the CNN model for the few-shot learning technique and helps improve the model performance. The CNN model is trained offline and for inference, in IoT devices, the model is taken as a pre-trained model. As the training happens offline, the few-shot learning-based CNN model doesn't incur any memory overhead in IoT devices. }

\subsection{Automatic Resource Estimation}

Execution, inference, and training of CNNs and DNNs for malware detection and other applications often incur a significant amount of resources. 
Deploying them on a single IoT device is not also always feasible due to the 
available limited resources. Furthermore, the on-going parallel execution of other applications on IoT devices such as sensing, and other computations 
minimize the available resources for CNN/DNN execution. 
Thus, the number of resources available in each node changes based on its workload. Instead of, manually calculating the parameters of CNN and estimating whether available resources on a node will be sufficient each time, a regression model is developed in this work. The binary regression model is trained using data such as CNN's parameters, memory requirements of these parameters, and available memory at each node. As output, the binary regression model gives an estimate of whether the CNN model inference can be performed on a single node or must be distributed onto multiple nodes. The rationale for adopting the binary regression is its low overhead and complexity along with higher efficiency.

As illustrated in Algorithm \ref{algo2}, the binary regression algorithm is constructed. The training features of the regressor are the parameters of CNN. So, the parameters of each layer are calculated. For each layer of CNN $\mathbb{C}$, (Line 3 - Line 5) the variables weight matrix $W$, bias $B$, and activation $A$ are collected and stored in the variable $var$. These variables contribute to parameter calculations of different layers in CNN (Line 6 - Line 11). As shown in Table \ref{tab1}, the input layer and pooling layer represented as $POOL$ of the Convolutional Neural Networks does not have any learnable parameters. So, parameters $par$ are zero for these two layers. For convolutional layer $CONV$, fully connected layer represented as $FC$ and softmax layer represented as $Softmax$, the parameters are calculated using the equation shown in Table \ref{tab1}.

\begin{algorithm}[htp!]
	\caption{Lightweight Linear Regression Algorithm}
	\label{algo2}
	\begin{algorithmic}[1] 
		
		\STATE {\textbf{Require}}: $B_{exe}$ (Benign application files), $M_{exe}$ (Malware application files)
		\STATE {\textbf{Input}}: $\mathfrak{Mem}[node], \mathbb{C}$
		
		
		\STATE \hskip 1em \textbf{define} $Regressor(\mathbb{C})$:
		
		\STATE \hskip 2em \textbf{for} {$layer \leftarrow \mathbb{C}$}: \textbf{do}
		\STATE \hskip 3em $ var \leftarrow f(W, B, A)$
		
		\STATE \hskip 3em \textbf{if} ${layer \rightarrow CONV}$
		\STATE \hskip 4em  $ par = f_{conv}(var)$
		\STATE \hskip 3em \textbf{elif} ${layer \rightarrow  (FC \vee Softmax)}$ 
		\STATE \hskip 4em $par = f_{FC}(var)$
		\STATE \hskip 3em \textbf{else}
		\STATE \hskip 4em $par = 0$
		\STATE \hskip 3em {\textbf{end if}}
		
		\STATE \hskip 3em $ \Bar(P).append(par)$ 
		
		\STATE \hskip 2em \textbf{end for}
		
		\STATE \hskip 2em $ \mathfrak{Mem}[model] \leftarrow N*batch\_size*\Bar(P)*1KB$ 
		\STATE \hskip 2em $X_R.features \leftarrow \{ W, A, B, \Bar(P), \mathfrak{Mem}[model],$
		\STATE \hskip 9em $\mathfrak{Mem}[node] \}$
		\STATE \hskip 2em $ Res \leftarrow \mathbb{R}:(X_R, \beta)$
		\STATE \hskip 2em \textbf{return} $Res$
		
	\end{algorithmic}
\end{algorithm}

If there are multiple convolutional and pooling layers, the parameters are calculated multiple times with different activation functions $A$. At last, the estimated parameters of each layer are appended to give $\bar(P)$ (Line 13 - Line 18). Then, the memory for the model $\mathfrak{Mem}[model]$, is calculated. The memory is a function of parameters $\bar(P)$ for each batch per N number of batches. It is assumed that each parameter needs one Kilo Byte (1KB) for inference, based on which the final memory required will be in MBs ($\sim 5MB$). $X_R.features$ represents the features to be given as input to the regressor $\mathbb{R}$ which predicts the resource estimations $Res$. The features in $X_R.features$ include, weight matrix $W$, bias $B$, activation $A$, parameters of CNN at each layer $\Bar(P)$, memory estimation of model $\mathfrak{Mem}[model]$ and memory available at each node $\mathfrak{Mem}[node]$. \journal{This resource estimation provides a prediction whether the inference can be performed on a single node or if it needs to be done using parallelism.}

\subsection{Workload- and resource-aware malware detection}



\journal{We develop a few-shot learning based convolutional neural network, trained on malware and benign samples. The inference task using this pre-tained malware detection model is partitioned and executed on different devices 
	\cite{Verbraeken'20}. It was also ensured that child devices have no access to complete information. 
	The partitioning is performed based on the independency of the nodes of the ML classifier, 
	represented as a graph, and the workload that could be accommodated on the parent and child devices \cite{Geng'19}. We provide an upper bound on the number of devices to which the task can be distributed.} 
As the ML architecture is defined during design time, the model parallelism and model splitting overheads do not affect during the runtime. 
The overhead to determine whether distributed ML is needed is minimal due to involved low-complex computations.

\begin{figure}
	\includegraphics[width=\linewidth]{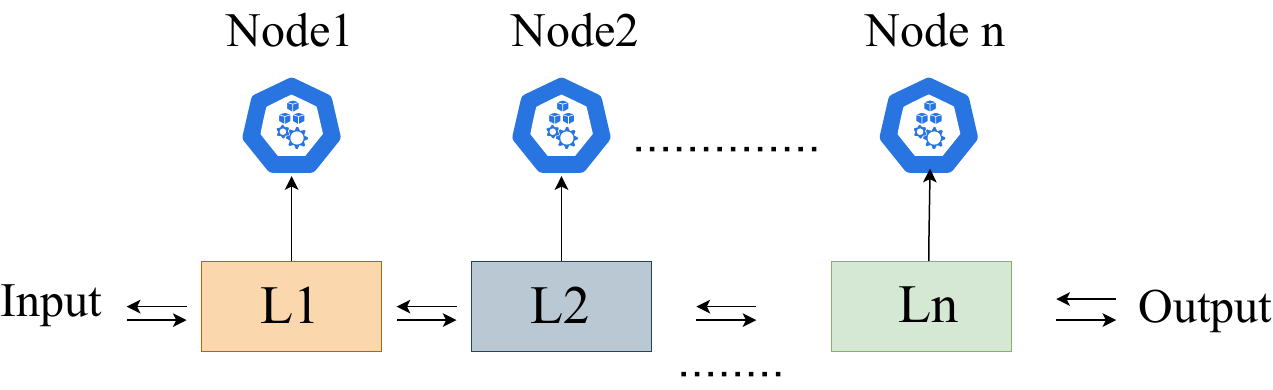}
	\caption{Model Distribution Over n Nodes}\label{fig:MP}
\end{figure}

Given the model is distributed on multiple IoT devices as shown in Figure \ref{fig:MP}, the 
accumulation of the gradients from the child nodes is a challenging task \cite{Verbraeken'20}. 
Techniques such as DistBelief \cite{Jeffrey'12} are highly dependent on the partitioning of the model. Thus, they can lead to varied performances in our case and hence not adaptable. 
We adapt AllReduce \cite{Pitch'09} paradigm in this project, where the parent node accumulates the gradients from the children nodes.
To update the gradients and perform other computations including inference, 
a synchronous Allreduce approach is utilized for better scalability \cite{Pitch'09}. 
However, a direct adaptation of such a method makes it vulnerable to faults such as the unavailability or garbage data from one device 
can stagnate or contaminate the whole process. 
To address such concerns, 
we deploy Downpour stochastic gradient descent (SGD) \cite{Duchi'11}. Downpour SGD is more resilient to machine failures and data manipulations, 
as it allows the training and inference to continue even if some model replicas are offline. 
It needs to be noted that the training happens offline, and inference is performed in real-time. 
To minimize the communication overheads, we let the parent device choose the child devices within a threshold radius $R$ for which the communication costs are lower and ensure the devices have a smaller workload to process. As frequent communication between parent and child nodes lead to large overheads,  
we let the system communicate whenever a device's output is required as input for another device.

\begin{algorithm}[htp!]
\caption{Pseudo-Code for Distributed Runtime Modelling of Malware Detection}
\label{algo3}
\begin{algorithmic}[1] 
	\STATE  {\textbf{Require}}: $M$ (Malware grayscale images), $B$ (Generated Benign grayscale images)
	\STATE  {\textbf{Input}}: $D^n = \{B + M\}, \mathfrak{Mem}[model] $

	
	\STATE \hskip 1em \textbf{define} $Distribute\_CNN\_model()$:
	\STATE \hskip 2em \textbf{for} {$ n \leftarrow range(0,x)$}: \textbf{do} 
	\STATE \hskip 3em \textbf{if} $\mathfrak{Mem}[model] \leq \mathfrak{Mem}[node]$
	\STATE \hskip 4em $ node.append(n)$
	\STATE \hskip 4em $ \mathfrak{Mem}[node] \leftarrow \mathfrak{Mem}[0] + \mathfrak{Mem}[1]+...+\mathfrak{Mem}[n]$
	
	\STATE \hskip 3em {\textbf{end if}}
	
	\STATE \hskip 2em\textbf{end for}
	
	\STATE \hskip 2em $l_1 = nn.layer1.cuda(0)$
	\STATE \hskip 2em $l_2 = nn.layer2.cuda(1) $
	\STATE \hskip 2em $\cdots$
	\STATE \hskip 2em $\cdots$
	\STATE \hskip 2em $l_n = nn.layern.cuda(n) $
	\STATE \hskip 2em $ model = nn.Sequential(l_1, l_2,...,l_n)$
	\STATE \hskip 2em $input = D^n.cuda(0) $
	\STATE \hskip 2em $ output = model(input) $
	
	\STATE \hskip 2em \textbf{return} $O_m$
\end{algorithmic}
\end{algorithm}

Algorithm \ref{algo3}, represents the Pseudo-code for proposed distributed runtime-based modeling of malware detection. The function to distribute CNN represented as, $Distribute\_CNN\_model()$, is called based on the output of the regressor. It also takes the memory estimation $\mathfrak{Mem}[model]$ of the CNN model for malware detection as input. It compares the model memory $\mathfrak{Mem}[model]$ and available memory at each node $\mathfrak{Mem}[node]$. It appends multiple node memory elements to find the number of nodes, required to distribute the model. The number of nodes $n$ should have a combined memory more than or equal to the model memory $\mathfrak{Mem}[model]$ (Line 3 - Line 5). If this condition is met, the CNN is distributed on $n$ nodes. The different layers of malware detection model $l_1, l_2,\cdots, l_n$ (Line 10 - Line 14) are divided on $n$ and trained. The input data is made available to the input layers, by passing them to the $node0$. Communication between the nodes is made possible due to the interdependent variables and back pass algorithm by the function $ model = nn.Sequential(l_1, l_2,...,l_n)$ (Line 15 - Line 17). 

\section{Results}
\label{results}

\subsection{Experimental Setup}


\tcad{For the IoT network setup, we deployed 20 IoT nodes encompassing Broadcom BCM2711, and quad-core Cortex-A72 (ARM v8) 64-bit boards. These nodes are connected through a wireless interface (WiFi). For the purpose of model parallelism, we deployed 
	multiple Jetson Nanos containing 128-core NVIDIA Maxwell architecture-based GPU and Quad-core ARM® A57 CPU. The 4 JetsonNano boards
	are deployed for employing model parallelism and providing access to 
	multiple CPU and GPU nodes to IoT nodes. 
	Each Jetson Nano board acts as a single entity. 
	We have obtained malware and benign applications from VirusTotal \cite{rvirus} with 12,500 benign samples and 70,000 malware samples that encompass 5 malware classes: backdoor, rootkit, trojan, virus, and worm. These files are executed in Sandbox to capture malware HPC attributes. These HPC attributes of benign and malware samples are converted to grayscale images of size 256 $\times$ 256. The benign and malware binary samples are also converted to grayscale images of size 256 $\times$ 256. In this image dataset, 70\% of the data is divided into the training set and 30\% of unseen data is taken as the test set. To further improve the model training and make it resilient to malware evolution, synthetic data generated using code-aware data generation technique based on few-shot learning technique is added to the training set. This synthetic data is also split into 70\%-30\% and used to augment the training and test data during the runtime. A CNN is built on all this data in offline and the inference task of test data is performed on multiple CPU and GPU nodes based on resource availability.}


\subsection{Simulation Results}

The inference is performed using a pre-trained convolution neural network algorithm. If the resources to perform inference are not enough, the malware detection task is off-loaded to multiple nodes. Table \ref{tabu3} represents the performance of different datasets when model parallelism is applied. These datasets have only a few samples and are populated with synthetic data generated using GANs. Difference models trained on HPC, binary, and combined datasets are divided over multiple nodes. We compare the performance of the proposed technique in terms of accuracy, F1-score, precision, and recall. \journal{We can observe the highest accuracy of 98.62\% in the case of training data containing both HPC and binary image data, and where model parallelism is performed on two nodes. This case performs well because there are more training samples in this case which improves the model learning capability. And also the inference model is only divided into two nodes, so the penalty is less compared to three or four-node model parallelism. The lowest accuracy of 89.45\% is observed in the case of the model trained on a binary dataset and performed model parallelism on four nodes. This is due to the limited features present in the binary dataset which help in detecting complex malware. And the penalty due to dividing the model into four nodes. With the increase in the number of nodes the model is divided, we can observe a minute penalty in performance.}

\begin{table}[htbp]
	\caption{Performance comparison of the proposed model with different MP algorithms}
	\vspace{-1em}
	\begin{center}
		\scalebox{0.8}
		{
			\begin{tabular}{|c|c|c|c|c|c|}
				\hline
				\textbf{Model} & \textbf{Nodes} & \textbf{Accuracy} & \textbf{Precision} & \textbf{Recall} & \textbf{F1-score} \\
				&  & \textbf{(\%)} & \textbf{(\%)} & \textbf{(\%)} & \textbf{(\%)} 
				\\
				\hline
				& 2 & 97.64  & 97.62 & 97.65 & 97.63 \\
				\cline{2-6}
				HPC & 3 & 97.24  & 97.24 & 97.26 & 76.45 \\ 
				\cline{2-6}
				(with MP) & 4 & 96.73 & 96.73 & 97.21 & 96.72 \\ 
				
				\hline
				& 2 & 91.62  & 91.63 & 91.62 & 91.70 \\
				\cline{2-6}
				Binares & 3 & 91.18  & 91.16 & 91.15 & 91.18 \\ 
				\cline{2-6}
				(with MP) & 4 & 89.45 & 89.43 & 89.42 & 89.45 \\ 
				
				\hline
				& 2 & 98.62  & 98.63 & 98.62 & 98.70 \\
				\cline{2-6}
				HPC + Binaries& 3 & 98.14  & 98.12 & 97.86 & 98.12 \\ 
				\cline{2-6}
				(with MP) & 4 & 97.12 & 97.13 & 97.12 & 97.12 \\ 
				
				\hline
				
			\end{tabular}
		}
		\label{tabu3}
	\end{center}
\end{table}

\subsection{Comparison with Previous Works}
Table \ref{tab4} presents the comparison of the proposed technique with the existing HPC-based malware detection techniques. We compare the performance of the proposed technique in terms of accuracy, F1-score, recall, latency, and area. All the models in table \ref{tab4} focus on malware detection based on HPC runtime features. Compared to the existing techniques the proposed CNN-based distributed training on HPC-based image data achieves the highest accuracy. It maintains an accuracy of 96.7\% while producing comparable latency and area values. 

\begin{table}[htb!]
\centering
\caption{Comparison with existing HPC-based detection techniques}
\label{tab3}
\scalebox{0.82}{
	\begin{tabular}{|c|c|c|c|c|c|}
		\hline
		
		\textbf{Model} &  \textbf{Accuracy} & \textbf{F1-score} & \textbf{Recall} & \textbf{Latency}  & \textbf{Area} \\
		& (\%) & (\%) & (\%) & (@ 10ns) &($\mu$m$^2$) \\
		\hline 
		OneR \cite{HPC_r1} & 81.00  & 81.00 & 82.00 & 1 & 1258\\
		\hline
		JRIP \cite{HPC_r1} & 83.00 & 83.00 & 84.00   & 4 & 1504\\
		\hline
		PART \cite{HPC_r1}& 81.00 & 81.50 & 83.10 & 6 & 2131\\
		\hline 
		J48 \cite{HPC_r1} & 82.00  & 82.00 & 82.00  & 9 & 1801\\
		\hline
		Adaptive-HMD \cite{adaptive_hmd} & 85.30 & 85.30 & 85.80 & 4 & 876 \\
		\hline
		SVM \cite{results_nn}& 73.90  & 73.60 & 77.20  & - & -\\
		\hline
		RF \cite{results_nn} & 83.50 & 83.40 & 82.20 & - & - \\
		\hline
		NN \cite{results_nn} & 81.10 & 81.10 & 81.60 & - & - \\
		\hline
		SMO \cite{results_smo} & 93.20 & 93.30 & 93.10 & 22  & 2466 \\
		\hline
		\textbf{Proposed} & \textbf{96.70} & \textbf{96.70} & \textbf{97.20} & \textbf{10} & \textbf{1044} \\
		\hline

\end{tabular}}
\label{tab4}
\vspace{1.8em}
\end{table}

The F1 score and recall also support the performance of the proposed technique compared to other techniques. From these results, it is evident that the proposed technique achieves state-of-the-art HPC-based malware detection accuracy. The latency is represented in terms of clock cycles of 10ns to measure inference time, obtained from the Synopsys DC tool. The inference time of a few tens of nanoseconds indicates real-time malware detection. 

\vspace{-0.5em}
\subsection{Impact on Latency with Proposed Technique}
\begin{figure}
\vspace{-6.0em}
\includegraphics[width=0.5\textwidth, height= 9cm]{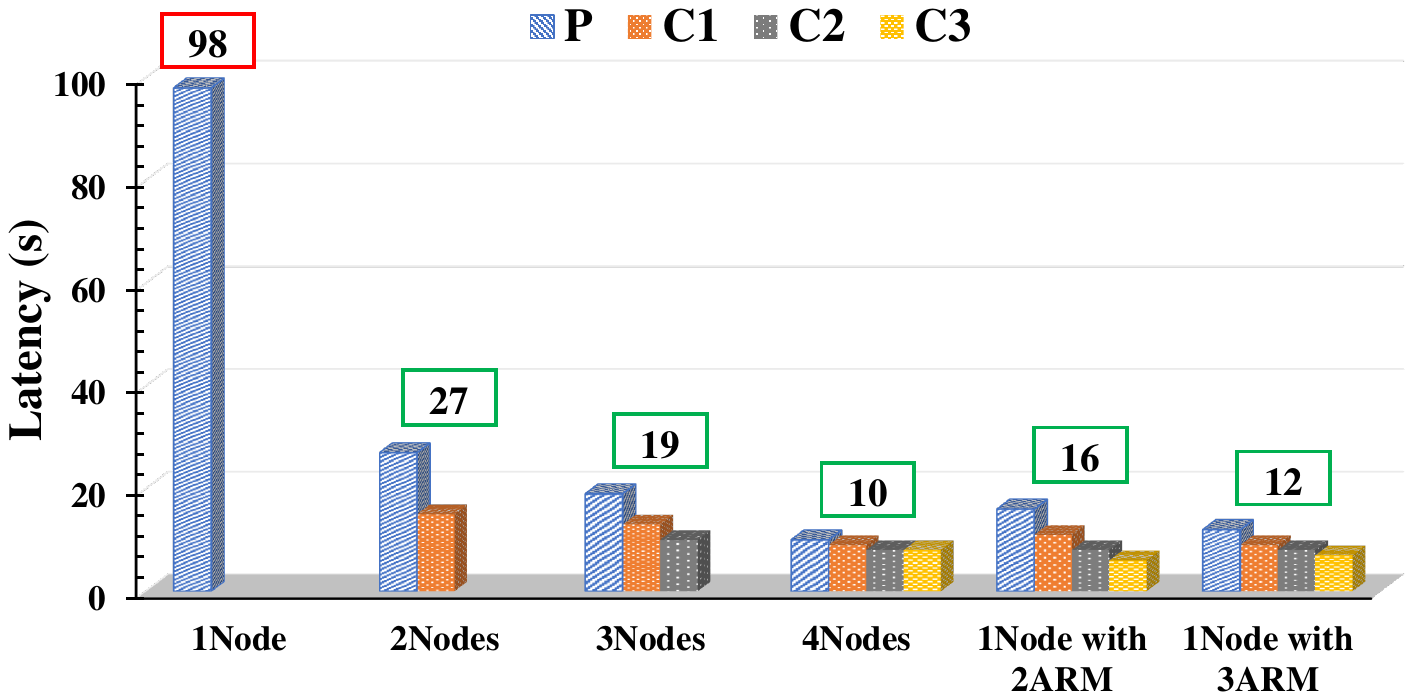}
\vspace{-7.5em}
\caption{Latency of Distributed learning for Malware Detection} 
\label{fig:latency}
\vspace{-1.5em}
\end{figure}

\vspace{-0.25em}
Normalized inference execution time is analyzed for cases a) the parent node has sufficient resources; b) the parent node does not have enough resources and outsources to multiple nodes. Figure \ref{fig:latency}, represent the latency of these cases. In Figure \ref{fig:latency}, Node represent a 128-core NVIDIA Maxwell architecture-based GPU present in Jetson Nano boards and the ARM represents the Quad-core ARM® A57 CPU present in Jetson boards. Also, P represents the parent node, C1 represents the first child node, C2 represents the second child node and C3 represents the third child node.
We observe that with an increase in the number of nodes the inference time decreases. As the parameters are divided over various nodes the execution time needed for inference decreases. 
As the executions run in parallel, the total latency to perform inference in model parallelism is the latency of node which takes the highest time to execute (usually the model parallelism latency is the latency of parent node P). In Figure \ref{fig:latency}, for the case of sufficient resources, it takes 98 seconds to perform the inference task. 
For the case of model parallelism, we can observe a speedup of 4 $\times$ when the inference task is parallelized between two nodes. If we further off-load the inference task to three nodes, an additional speedup of about 1.5$\times$ is observed. 
The ARM boards used as child nodes in model parallelism also produced notable speed-ups. We observed an overall speedup of 9.8$\times$ while using four Jetson Nano boards. 

\subsection{Impact of Proposed Technique on Resource Consumption}
\begin{figure}
\vspace{-1.0em}
\includegraphics[width=\linewidth]{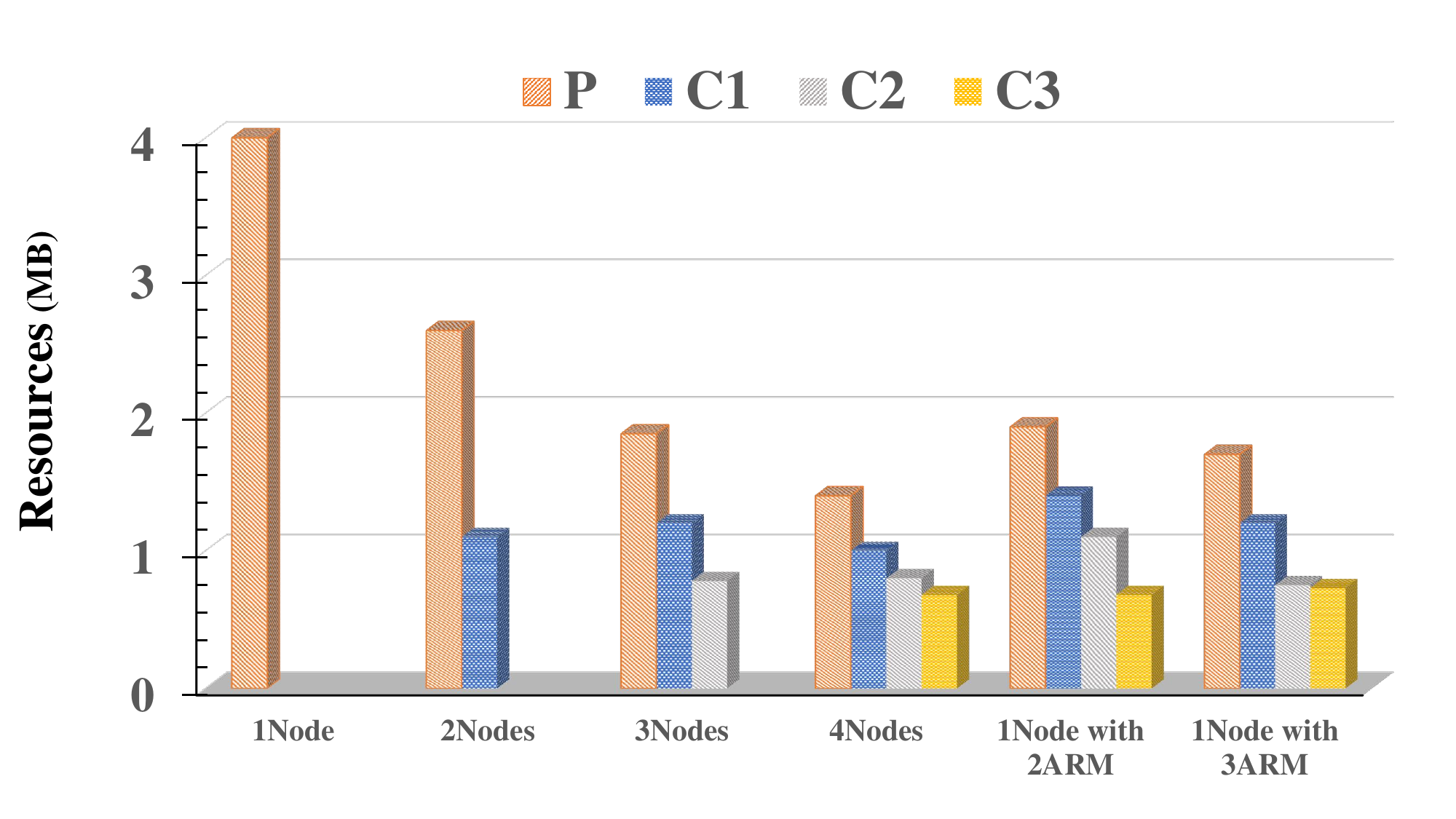}
\vspace{-2.5em}
\caption{Resource Consumption for Inference Over n Nodes} 
\label{fig:resources}
\vspace{-1.5em}
\end{figure}

\revision{The resource consumption of different worker nodes can be observed in Figure \ref{fig:resources}. 
In Figure \ref{fig:resources}, P represents the parent node, C1 represents the first child node, C2 represents the second child node and C3 represents the third child node. 
We observe the resource consumption for the following scenarios: a) the parent node has sufficient resources; b) the parent node does not have enough resources and outsources to multiple nodes.
The inference task takes 4 MB of data to complete. 
In the first case, the single parent node P can provide this data to complete the inference task. In other cases, the inference task is divided between multiple nodes (model parallelism), so the data required is also divided into multiple nodes. 
We can observe that the resources are not equally consumed in the parent and child nodes. This is because the parent node usually has additional steps to perform, like the gradient collection from child nodes, adding them, etc, so, it consumes high resources. 
When compared to using a single node, in model parallelism, the required resources are provided from various nodes, and this helps to improve the processing speed.} 

\vspace{-1.5em}

\subsection{ASIC Implementation of Proposed Technique}

We present the hardware implementation of the classifiers on ASIC to estimate resource utilization. 
The power, area, and energy values are reported at 100MHz. 
We used Design Compiler Graphical by Synopsys to obtain the area for the models. Power consumption is obtained using Synopsys Primetime PX. The post-layout area, power, and energy are summarized in Table \ref{tab:harm_res}. The resource utilization of the binary regression model is significantly less, whereas the CNN consumes high power, energy, and area on-chip ( Table \ref{tab:harm_res} ), hence we split the CNN model across the neighboring nodes with the available resource for inference computation. The post-layout energy numbers were almost $\approx$ 1.8 $\times$ higher than the post-synthesis results. This increase in energy is mainly because of metal routing resulting in layout parasitics. As the tool uses different routing optimizations, the power, area, and energy values keep changing with the classifiers' composition and architecture.

\begin{table}[htb!]
\centering
\caption{Post synthesis hardware results of the classifiers (@100MHz) when deployed }\label{tab:harm_res}
\scalebox{0.82}{
\begin{tabular}{|c|c|c|c|}
\hline

Model &  Power ($mW$) & Energy ($mJ$) & Area ($mm^2$) \\
\hline 
CNN & 82.45  & 5.12 & 4.42\\
\hline
Regressor & 27.81 & 2.52 & 1.18 \\	
\hline
\end{tabular}}
\vspace{-1.8em}
\end{table}

\vspace{-0.1em}

\section{Conclusion}
\label{conclusion}

With the proposed resource- and workload-aware model parallelism-based malware detection technique employs distributed training to enable better security for resource-constrained IoT devices. The metrics of distributed training on multiple nodes are analyzed and a speed-up of 9.8$ \times$ is observed compared to on-device training. From the results presented, it is also evident that the proposed technique produces state-of-the-art malware detection accuracy of 96.7\% among HPC-based detection techniques. We also furnished the ASIC implementations of the CNN classifier trained using the proposed technique and the lightweight logistic regressor trained to classify the availability of resources. Thus, the proposed technique is reliable and accurate for malware detection in IoT devices.

	\bibliographystyle{IEEEtran}
	\bibliography{reference.bib}
	
	
	\newpage
	
	\begin{IEEEbiography}[{\includegraphics[width=1in,height=1.25in,clip,keepaspectratio]{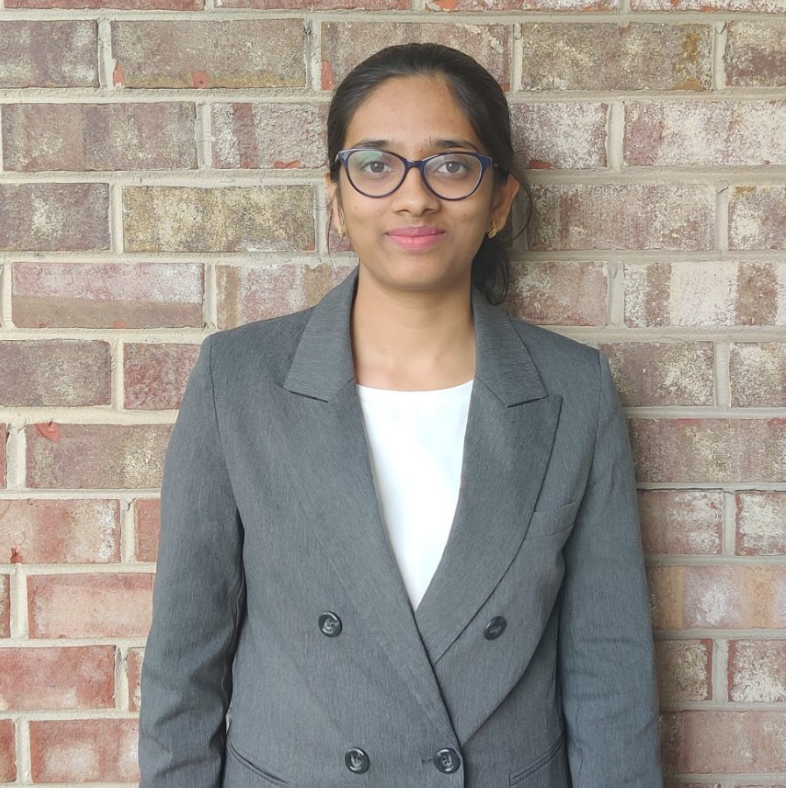}}]{Sreenitha Kasarapu}
		is a Ph.D. student, currently conducting her research under the supervision of Dr. Sai Manoj P D, an Assistant professor at the Electrical and Computer Engineering Department, George Mason University, Fairfax, VA, USA. She previously worked as a research assistant at GMU. Sreenitha's present research interest includes IoT network security, computer vision, image processing, and time series analysis. She published her work in AICAS'20 and actively participated in research projects for malware detection. She received her Bachelor in Technology degree in Electronics and Communication Engineering from Jawaharlal Nehru Technological University Hyderabad (JNTUH), Hyderabad, India in 2019. At that time, she won second prize at National Level Technical Symposium held by the Indian Society for Technical Education (ISTE) for her paper on sensor technology and participated in several peer-reviewed paper presentations.

	\end{IEEEbiography}

	\begin{IEEEbiography}[{\includegraphics[width=1.05in,height=1.5in,clip,keepaspectratio]{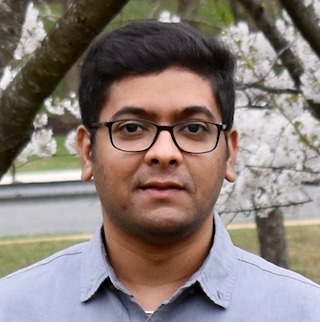}}]{Sanket Shukla}
		received his bachelor’s degree in Electronics Engineering, in 2015, from Mumbai University and his master’s degree (M.Sc.) in Computer Engineering, in 2021, from George Mason University. He is currently pursuing a Ph.D. degree with the Electrical and Computer Engineering department, George Mason University under the supervision of Dr. Sai Manoj PD. He conducts research in developing machine learning and deep learning-based solutions for IoT and cybersecurity. He has published research papers in DATE, DAC, ICCD, GLSVLSI, ICMLA and ICTAI conferences and have also reviewed several journals and research papers. His research work submitted to DATE 2023 was recognized and nominated for the best paper award. His research interests include malware detection, cybersecurity, federated learning, energy and computational efficient machine learning and deep learning for security on IoTs and computer systems.
		
	\end{IEEEbiography}

	\begin{IEEEbiography}[{\includegraphics[width=1.2in,height=1.0 in,clip,keepaspectratio]{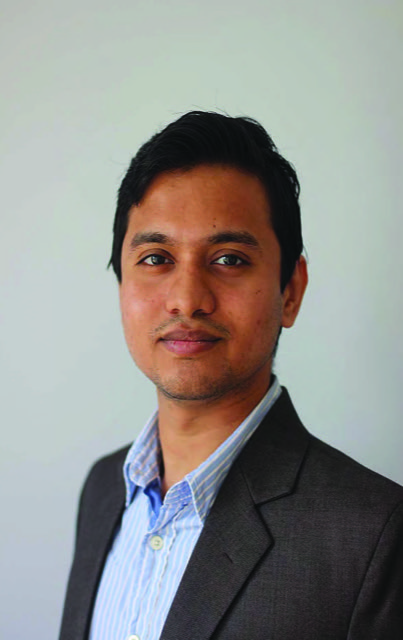}}]{Sai Manoj P D}
		(S'13-M’15) is an assistant professor at George Mason University. Prior joining to George Mason University (GMU) as an assistant professor, he served as research assistant professor and post-doctoral research fellow at GMU and was a post-doctoral research scientist at the System-on-Chip group, Institute of Computer Technology, Vienna University of Technology (TU Wien), Austria. He received his Ph.D. in Electrical and Electronics Engineering from Nanyang Technological University, Singapore in 2015. He received his Masters in Information Technology from International Institute of Information Technology Bangalore (IIITB), Bangalore, India in 2012.
		His research interests include on-chip hardware security, neuromorphic computing, adversarial machine learning, self-aware SoC design, image processing and time-series analysis, emerging memory devices and heterogeneous integration techniques. He won best paper award in Int. Conf. On Data Mining 2019, and his works were nominated for best paper award in prestigious conferences such as Design Automation \& Test in Europe (DATE) 2018, International Conference on Consumer Electronics 2020, and won Xilinx open hardware contest in 2017 (student category). He is the recipient of the ``A. Richard Newton Young Research Fellow'' award in Design Automation Conference, 2013.
	\end{IEEEbiography}

\end{document}